\documentclass[journal,10pt]{IEEEtran}
\pdfoutput=1
\usepackage{cite}

\usepackage{amsmath}
\usepackage{algorithm}
\usepackage{algorithmic}
\usepackage{subfigure}
\usepackage{amssymb}
\usepackage{graphicx}
\usepackage{empheq}

\usepackage{array}
\usepackage{mathtools}
\usepackage{graphicx}
\usepackage{cases}
\usepackage[colorinlistoftodos]{todonotes}
\usepackage[colorlinks=true, linkcolor=black,citecolor=black,urlcolor=black]{hyperref}
\usepackage{indentfirst}
\usepackage{multirow}
\usepackage{booktabs}
\usepackage{makecell}
\allowdisplaybreaks[4]

\usepackage{url}
\usepackage[hyphenbreaks]{breakurl}

\begin{document}

% \linenumbers

\title{Swarm of Robotic Aerial Base Stations for mmWave Multi-Hop Backhauling}

\author{
Yuan~Liao, \IEEEmembership{Student Member,~IEEE}, Vasilis~Friderikos, \IEEEmembership{Member,~IEEE}, Halim Yanikomeroglu,  \IEEEmembership{Fellow,~IEEE}

\thanks{Yuan Liao and Vasilis Friderikos are with the Department of Engineering, King's College London, London WC2R 2LS, U.K. (e-mail: yuan.liao@kcl.ac.uk; vasilis.friderikos@kcl.ac.uk).

Halim Yanikomeroglu is with the Non-Terrestrial 
Networks (NTN) Lab, Department of Systems and Computer Engineering, Carleton University, Ottawa, ON K1S 5B6, Canada (e-mail: halim@sce.carleton.ca).}
}


\maketitle

\begin{abstract}

Robotic aerial base stations (RABSs) that are able to anchor at tall urban landforms are expected to bring further flexibility to millimeter-wave (mmWave) multi-hop backhaul networks in highly dense urban environments. In this paper, a swarm of RABSs are deployed to construct a dynamic mmWave backhaul network according to the traffic spatial distribution, and relocate their positions in subsequent time epochs according to the traffic temporal dynamic. The overall energy efficiency of the proposed framework is maximized by determining the RABS deployment, relocation and route formation under the channel capacity and hop constraints. The problem is formulated as a mixed-integer linear fractional programming (MILFP) and a two-stage method is developed to overcome the computational complexity. A wide set of numerical investigations reveal that compared to fixed small cells, only half as many RABSs are required to cover the same volume of traffic demand.  
\end{abstract}


\IEEEpeerreviewmaketitle

\section{Introduction}
\label{introduction}

Ultra-dense networks are inevitably required to satisfy the rapidly growing traffic demand and the increasingly heterogeneous distribution in 6G cellular networks. To overcome the augmented capital expenditures, we employ a swarm of robotic aerial base stations (RABSs) as a flexible and cost-effective solution for network densification. Taking inspiration from robotic grippers, RABSs are a novel prototype of aerial base stations that feature a dexterous anchoring mechanism installed on conventional unmanned aerial vehicle (UAV) base stations. Hence, they can attach autonomously on lampposts when providing wireless service, and fly to other grasping positions via controllable maneuverability to adapt to the traffic dynamics \cite{friderikos2021airborne,liao2023optimal,liao2023robust}. However, unlike terrestrial base stations connecting with the core network through high-capacity fibre links, RABSs necessitate wireless backhaul owing to their frequent mobility. In light of this, in this letter, we employ millimeter-wave (mmWave) frequencies for backhaul links due to their increased capacity and faster transmission. Additionally, we establish multi-hop routes to address the critical concern that mmWave channels are more susceptible to obstruction by obstacles.

To satisfy the rapidly growing traffic demand in mobile networks, the use of mmWave spectrum is expected to offer high-throughput wireless access. The authors of \cite{lun2016millimetre} verify the feasibility of deploying fixed mmWave ultra-dense outdoor small cells installed on street-level fixtures. The work \cite{palizban2017automation} utilizes the wall-mounted mmWave base station to provide line-of-sight (LoS) coverage in urban environment. Moreover, the benefits of mmWave backhaul come at the cost of considerably higher propagation losses, resulting in a reduced transmission range compared to lower frequencies. This limitation can be addressed by establishing a multi-hop topology using relay nodes. In \cite{mcmenamy2019hop}, by deploying small cells densely on a Manhattan-type geometry, the network flow of the backhaul network is maximized by route formation. The resource allocation for multi-hop backhaul networks is optimized according to the traffic demands in \cite{pu2018traffic,semiari2017inter}. Furthermore, to address the fluctuations in dynamic traffic, several studies employ UAVs as movable base stations to perform wireless backhaul tasks. For instance, the authors of \cite{gapeyenko2018flexible} measure the system performance when considering both heterogeneous mobility of blockers and UAVs. The utility function of an UAV-assisted multi-hop backhaul network is maximized in \cite{challita2017network} and the authors of \cite{qiu2019joint} study the deployment strategy for UAVs.

\begin{figure}[!t]
\centering
\setlength{\abovecaptionskip}{-0.1cm}
\includegraphics[width=0.85\linewidth]{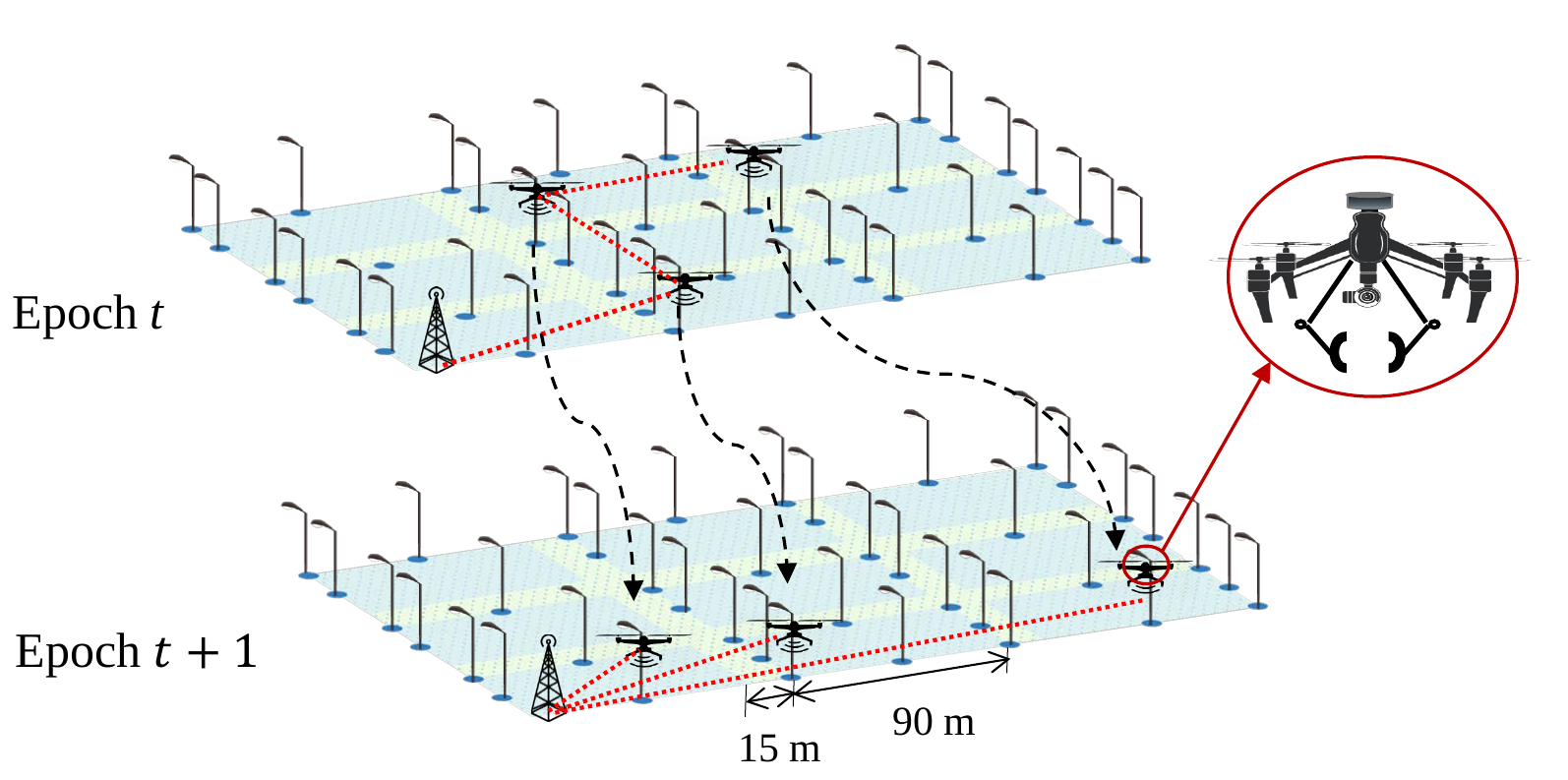}
\caption{A mmWave hackhaul network constructed by a swarm of RABSs: blue squares represent the building obstacles, the small blue circles are the candidate locations, the red and black lines represent the backhaul channels and the flying trajectory when relocating.}
\label{Toy_example}
\vspace{-0.65cm}
\end{figure}

In this letter, we present a flexible mmWave multi-hop backhaul network operated by a swarm of RABSs. Our approach is inspired by and developed from the work in \cite{mcmenamy2019hop}. This paper mainly demonstrates how movable RABSs can enhance the system performance, especially in a heterogeneous and dynamic environment. Specifically, the contribution of this letter is summerized as follows. Firstly, to overcome the endurance issue of the conventional UAV base stations studied in \cite{gapeyenko2018flexible,challita2017network,qiu2019joint}, we propose a flexible backhaul network configuration operated by a novel aerial base station prototype, i.e., RABS, which combines both flying and grasping capabilities. It could prolong the service time because the grasping power is much lower than the hovering/flying power of conventional hovering-based UAV base stations (tens of Watts versus hundreds of Watts) \cite{liao2023optimal,zeng2019energy}. Secondly, instead of the uniform traffic distribution assumed in \cite{mcmenamy2019hop,semiari2017inter}, we focus on the heterogeneous distribution and dynamic changes of traffic demand in spatial and temporal domains \cite{wang2015approach}. Furthermore, thanks to their inherent grasping and flying capabilities, RABSs can establish a backhaul network while anchoring on lampposts and relocate their grasping positions to adapt to traffic dynamic. Simulation results demonstrate that only half the number of RABSs is required to meet the same volume of traffic demands compared to densely-deployed small base stations. Thirdly, a mixed-integer linear fractional programming (MILFP) is formulated to maximize energy efficiency, defined as the ratio of achievable network flow to energy consumption, by determining the optimal deployment, movement, and multi-hop routing of RABSs. A two-stage method, which includes the column generation technique, the truncated Bellman-Ford algorithm, and the total unimodularity analysis, is proposed to efficiently solve the problem. Simulation results demonstrate that the proposed method achieves a favorable trade-off between high solution quality and low computational complexity.

\section{System Model and problem formulation}
\label{systemmodel}

The set of all candidate locations for RABSs grasping is denoted as $\mathcal{V}$. A Manhattan-type map is characterized by a graph $\mathcal{G} = (\mathcal{V}^a,\mathcal{E})$, where $\mathcal{V}^a = \{0\} \cup \mathcal{V}$ and $0$ is the index of the macro base station (MBS), $\mathcal{E}$ is the set of edges, which denotes the LoS links in mmWave backhaul networks. It should be noted that, similar to \cite{palizban2017automation,lun2016millimetre,mcmenamy2019hop}, this work only considers the unobstructed LoS channel for mmWave channel, while connections cannot be successfully established when obstructed by buildings, as mmWave frequencies are highly prone to obstruction by obstacles. In other words, in Fig. \ref{Toy_example}, RABSs deployed in the middle of a street have links along and across the street, while RABSs on the corners have diagonal links. Accordingly, the rate capacity of edge $(i,j) \! \in \! \mathcal{E}$ is calculated by \cite{mcmenamy2019hop},
\begin{equation}
\begin{aligned}
\label{rate_capacity}
R_{(i,j)} = B \min [ \log_2 (1+10^{0.1\times(\textsf{SNR}_{(i,j)} - 3)}), \textsf{SE}_{\textrm{max}}],
\end{aligned} 
\end{equation}
where $B$ is the available bandwidth, $\textsf{SE}_{\textrm{max}}$ is the maximum spectral efficiency in bps/Hz, and $\textsf{SNR}_{(i,j)}$ is the signal-to-noise ratio in dB. As the ultra-dense street-level networks studied in \cite{lun2016millimetre}, the interference between different backhaul links can be significantly eliminated when mmWave antennas generate highly directional narrow beams, making the network noise-limited rather than interference-limited. Therefore, in this paper, we suppose that all backhaul links share the same spectrum pool to enhance spectral efficiency. Besides, the entire serving time of RABSs is divided into $T$ epochs with an equal duration denoted by $\eta$. Although the traffic distribution in urban regions shows high inhomogeneity in both spatial and temporal domains \cite{wang2015approach}, it can still be predicted accurately from the typical experienced data via several existing methods, such as machine learning techniques \cite{feng2018deeptp}. Therefore, similar to \cite{liao2023optimal,liao2023robust}, the traffic pattern is assumed to be known and unchanging during each epoch. During each epoch, there are $N$ RABSs available to construct a multi-hop backhaul network to relay traffic to the MBS based on the spatial traffic distribution. When the traffic pattern changes in the subsequent time epoch, RABSs could relocate to establish a new network topology. The application scenario is depicted in Fig. \ref{Toy_example}.

As noted in \cite{mcmenamy2019hop}, whilst extending the coverage radius, multi-hop backhauling brings an additional challenge of the increased latency for every hop. To satisfy the latency requirements, the maximum allowed hop needs to be limited and is denoted by $H$. We traverse all feasible routes within the graph $\mathcal{G}$ satisfying the hop constraints $H$. These routes start from a candidate location within the set $\mathcal{V}$ and end at the MBS, and we refer to this set of routes as $\mathcal{P}$. To formulate the problem, we introduce three sets of variables as follows. The binary variable $x^t_i \in \{0,1\}$ indicates if the candidate location $i$ is selected to deploy a RABS at epoch $t$. The variable $y_{(i,j)}^{(t-1,t)}$ represents whether a RABS will fly from the lamppost $i$ to $j$ between two adjacent epochs, $t-1$ and $t$. The continuous variable $f^t_p$ denotes the volume of traffic flow in the route $p \in \mathcal{P}$ at epoch $t$. Accordingly, the energy efficiency metric is defined as follows:
\begin{equation}
\begin{aligned}
\label{energy_efficiency}
\textsf{EE} = \frac{\sum_{t=1}^{T}\sum_{p \in \mathcal{P}} \eta f^t_p }{\sum_{t=0}^{T} \sum_{(i,j) \in \mathcal{E}} E^{\textrm{fly}}_{(i,j)}y_{(i,j)}^{(t-1,t)} + NT(E^t + E^g) }, 
\end{aligned} 
\end{equation}
where $E^{\textrm{fly}}_{(i,j)}$ denotes the propulsion energy consumed by a RABS flying from the candidate location $i$ to $j$, while $E^t$ and $E^g$ denote the transmission and grasping energy, respectively\cite{liao2023optimal}. It can be observed that the numerator and denominator of \eqref{energy_efficiency} calculate the achievable network flow in bits and the total energy consumption in Joule, respectively. Furthermore, there are three sets of constraints used to formulate the problem.

\textit{1) Capacity constraints for mmWave channels:} Initially, we define a subset $\mathcal{P}_{(i,j)} \subseteq \mathcal{P}$ denoting all feasible routes passing the edge $(i,j)$. We then have the following constraints:
\begin{subequations}
\label{capacity_con}
\begin{empheq}[left={\empheqlbrace\,}]{align}
& \sum_{p \in \mathcal{P}_{(i,j)}} f^t_p \leq x^t_i  R_{(i,j)}, \; \forall (i,j) \in \mathcal{E}, \, \forall t \in \{1,...,T\}, \label{capacity_con_i} \\
& \sum_{p \in \mathcal{P}_{(i,j)}} f^t_p \leq x^t_j R_{(i,j)}, \; \forall (i,j) \in \mathcal{E}, \, \forall t \in \{1,...,T\}. \label{capacity_con_j} 
\end{empheq}
\end{subequations}
The right hand side of \eqref{capacity_con} indicates that only when both candidate locations $i$ and $j$ are deployed with RABSs, the traffic can traverse the edge $(i,j)$, and the accumulated flows, calculated by the left hand side of \eqref{capacity_con}, would not exceed the maximum value $R_{(i,j)}$. 

\textit{2) Traffic demand constraints:} The following constraints guarantee that all flows sourcing from the candidate location $i$ would not exceed its traffic demand:
\begin{equation}
\begin{aligned}
\label{demand_con}
\sum_{p \in \mathcal{P}_i} f^t_p \leq x^t_i D^t_i, \; \forall i \in \mathcal{V}, \, \forall t \in \{1,...,T\}, 
\end{aligned} 
\end{equation}
where $\mathcal{P}_i$ is a subset of $\mathcal{P}$ indicating all feasible routes sourcing from the candidate location $i$. 

\textit{3) Degree constraints for RABSs relocating and deployment:} Similar as in \cite{liao2023optimal}, we have the following degree constraints to link the RABSs relocating process and deployment:
\begin{subequations}
\label{degree_con}
\begin{empheq}[left={\empheqlbrace\,}]{align}
& \sum_{i \in \mathcal{V}} y_{(i,j)}^{(t,t+1)} = x^t_i, \; \forall i \in \mathcal{V}, \, \forall t \in \{0,1,...,T-1\}, \label{degreein_con} \\
& \sum_{i \in \mathcal{V}} y_{(i,j)}^{(t-1,t)} = x^t_j, \; \forall j \in \mathcal{V}, \, \forall t \in \{1,2,...,T\}, \label{degreeout_con} 
\end{empheq}
\end{subequations}where \eqref{degreein_con} depicts whether the RABS departs from the candidate location $i$ or not, while \eqref{degreeout_con} indicates the landing process.

Based on the above preliminaries, the energy efficiency maximization problem can be formulated as follows:
\begin{subequations}
\label{formulation}
\begin{align}
\; & \max_{\{x^t_i\},\{y_{(i,j)}^{(t-1,t)}\},\{f^t_p\}} \textsf{EE} \label{Obj} \\
s.t.
\; & \eqref{capacity_con},\, \eqref{demand_con}, \,\eqref{degree_con}, \\
\; & \sum_{i \in \mathcal{V}} x^t_i \leq N, \; \forall t \in \{1,2,...,T\}, \label{RABSnumber_con} \\
\; & x^t_i \in \{0,1\}, \; \forall i \in \mathcal{V}, \, \forall t \in \{1,2,...,T\},  \\
\; & y_{(i,j)}^{(t-1,t)} \in \{0,1\}, \; \forall (i,j) \in \mathcal{E}, \, \forall t \in \{1,2,...,T\}, \label{yvar_bin}  \\
\; & f^t_p \geq 0, \; \forall p \in \mathcal{P}, \, \forall t \in \{1,2,...,T\},
\end{align}
\end{subequations}where the constraints \eqref{RABSnumber_con} indicate that there are at most $N$ RABSs that can be deployed. It can be observed that the problem \eqref{formulation} is a MILFP. Although Lemma 4 in \cite{you2009dinkelbach} proves that the MILFP can be solved to optimality via Dinkelbach's algorithm, solving \eqref{formulation} is still challenging for the following two main reasons. First, a mixed integer linear programming needs to be solved at each iteration of the Dinkelbach's algorithm, which is still NP-hard. Second, the scale of the problem \eqref{formulation} grows sharply because the cardinality of the set $\mathcal{P}$, denoted by $|\mathcal{P}|$, increases exponentially as $H$ and $|\mathcal{V}|$ increase. To this end, a two-stage method is developed in the following Section \ref{proposed_method} to overcome the curse of dimensionality. 

\vspace{-0.2cm}
\section{The Proposed Two-stage Method}
\label{proposed_method}

The problem \eqref{formulation} is decoupled into two sub-problems and solved separately. Firstly, the traffic flow on the backhaul network is maximized via column generation and the truncated Bellman-Ford algorithm, and then the energy consumption is minimized through linear programming by exploring the total unimodularity structure of the problem. 

\subsubsection{Maximize the traffic flow via column generation}
\label{ColumnGeneration}

Hereafter we aim to maximize the numerator of \eqref{energy_efficiency}, i.e., the accumulated traffic flow of the backhaul network. Observe that when only the traffic flow is considered, the problem can be decoupled into $T$ sub-problems for each epoch and solved separately. Numerically, the network flow maximization problem for epoch $t$ can be simplified from \eqref{formulation} as
\begin{subequations}
\label{flow_max_form}
\begin{align}
\; & \max_{\{x^t_i|\forall i \in \mathcal{V} \},\{f^t_p| \forall p \in \mathcal{P}\}} \sum_{p \in \mathcal{P}} \eta f^t_p \label{flow_max_obj} \\
s.t.
\; & \sum_{p \in \mathcal{P}_{(i,j)}} f^t_p \leq x^t_i  R_{(i,j)}, \; \forall (i,j) \in \mathcal{E}, \label{flow_max_capacity_con_i} \\
\; & \sum_{p \in \mathcal{P}_{(i,j)}} f^t_p \leq x^t_j R_{(i,j)}, \; \forall (i,j) \in \mathcal{E}, \label{flow_max_capacity_con_j} \\
\; & \sum_{p \in \mathcal{P}_i} f^t_p \leq x^t_i D^t_i, \; \forall i \in \mathcal{V},  \label{flow_max_demand_con} \\
\; & \sum_{i \in \mathcal{V}} x^t_i \leq N,  \label{flow_max_RABSnumber_con} \\
\; & x^t_i \in \{0,1\}, \; \forall i \in \mathcal{V}, \label{flow_max_integer_con}  \\
\; & f^t_p \geq 0, \; \forall p \in \mathcal{P}.
\end{align}
\end{subequations}
As illustrated above, the main challenge for solving \eqref{flow_max_form} is the extremely large size of the set $\mathcal{P}$. Given this consideration, we employ the column generation method to solve this large-scale problem.

Column generation is an efficient method for solving large-scale linear programming problems \cite{lubbecke2005selected}. The principal idea is to activate a small subset of variables and solve the reduced scale problem, then repeatedly add the variables that can improve the objective function to this subset until no more such variables can be found. In the following, column generation is employed to solve the linear relaxation of \eqref{flow_max_form}, denoted as (\ref{flow_max_form}-LR) for ease of notation. We first select and activate a subset $\mathcal{P}' \subseteq \mathcal{P}$, and relax the binary variables in \eqref{flow_max_integer_con} into continuous variables. This problem is referred as the restricted master problem for (\ref{flow_max_form}-LR) and can be written as
\begin{subequations}
\label{master_form}
\begin{align}
\; & \max_{\{x^t_i|\forall i \in \mathcal{V} \},\{f^t_p| \forall p \in \mathcal{P}'\}} \sum_{p \in \mathcal{P}'} \eta f^t_p \label{master_obj} \\
s.t.
\; & \sum_{p \in \mathcal{P}'_{(i,j)}} f^t_p \leq x^t_i  R_{(i,j)}, \; \forall (i,j) \in \mathcal{E}, \label{master_capacity_con_i} \\
\; & \sum_{p \in \mathcal{P}'_{(i,j)}} f^t_p \leq x^t_j R_{(i,j)}, \; \forall (i,j) \in \mathcal{E}, \label{master_capacity_con_j} \\
\; & \sum_{p \in \mathcal{P}'_i} f^t_p \leq x^t_i D^t_i, \; \forall i \in \mathcal{V},  \label{master_demand_con} \\
\; & \sum_{i \in \mathcal{V}} x^t_i \leq N,  \label{master_RABSnumber_con} \\
\; & 0 \leq x^t_i \leq 1, \; \forall i \in \mathcal{V}, \label{master_integer_con}  \\
\; & f^t_p \geq 0, \; \forall p \in \mathcal{P}',
\end{align}
\end{subequations}
where $\mathcal{P}'_{(i,j)}$ and $\mathcal{P}'_i$ is the activated subset corresponding to $\mathcal{P}_{(i,j)}$ and $\mathcal{P}_i$, respectively. Problem \eqref{master_form} is a linear programming and we can write its dual as
\begin{subequations}
\label{pricing_form}
\begin{align}
\; & \max_{\{ \alpha_{ij} \},\{ \beta_{ij} \}, \atop \{\gamma_i\}, \{\delta \}, \{\zeta_i\}} N \delta + \sum_{i \in \mathcal{V}} \zeta_i \label{pricing_obj} \\
s.t.
\; & \sum_{(i,j) \in p} (\alpha_{ij} + \beta_{ij}) + \gamma_{s_p} \geq \eta, \; \forall p \in \mathcal{P}', \label{pricing_con_f} \\
\; & \sum_{j \in \mathcal{V}}(R_{(i,j)}\alpha_ij \! + \! R_{(j,i)}\beta_{ji}) \! + \! D^t_i \gamma_i \leq \delta \! + \! \zeta_i,  \forall  i \! \in \! \mathcal{V}, \label{pricing_con_x} \\
\; & \alpha_{ij} \geq 0, \, \beta_{ij}\geq 0, \, \gamma_i\geq 0, \, \delta\geq 0, \, \zeta_i\geq 0, \; \forall i, j, \label{pricing_nonnegative}
\end{align}
\end{subequations}where $\{ \alpha_{ij} \}, \{ \beta_{ij} \}, \{\gamma_i\}, \, \{\delta \}$ and $ \{\zeta_i\}$ are dual variables corresponding to the constraints in \eqref{master_form}, $(i,j) \in p$ indicates that the route $p$ passes through the edge $(i,j)$, and $s_p \in \mathcal{V}$ denotes the source node of route $p$. Problem \eqref{pricing_form} is normally referred as the pricing problem. According to the nominal principle of column generation, we have the following Lemma 1. 

\textit{\textbf{Lemma 1:}} If the constraints \eqref{pricing_con_f} are even satisfied for all $p \!\in \! \mathcal{P}$ (i.e., not only for $p \! \in \! \mathcal{P}'$), the result solved from the problem \eqref{master_form} with the restricted subset $\mathcal{P}'$ is optimal for (\ref{flow_max_form}-LR).

\textit{Proof:} Please refer to Section 2.1 of \cite{lubbecke2005selected}. $\square$

Lemma 1 illustrates that if there exists a route $p \! \in \! \mathcal{P}$ violating the constraints \eqref{pricing_con_f}, the objective value \eqref{master_obj} can be further improved by adding this route to $\mathcal{P}'$ and re-solve the problem \eqref{master_form}. Otherwise the optimal solution of (\ref{flow_max_form}-LR) can be achieved with the restricted route set $\mathcal{P}'$. 

However, an unresolved challenge remains after Lemma 1 in determining whether all $p \!\in \! \mathcal{P}$ satisfy the constraints \eqref{pricing_con_f}, or if we can find a feasible route $p$ that violates \eqref{pricing_con_f}. We then illustrate that this problem can be efficiently solved by considering it as a hop-constrained shortest path problem. By analyzing the constraints \eqref{pricing_con_f}, we can assign the weight $(\alpha_{ij} \! + \! \beta_{ij})$ to the corresponding edge $(i,j) \! \in \! \mathcal{E}$, and find the shortest path from the MBS to the candidate location $s \! \in \! \mathcal{V}$ with maximum $H$ hops. If the shortest path is greater than or equal to $(\eta \! - \! \gamma_s)$, then there is no feasible route sourcing from $s$ and hence violates the constraints \eqref{pricing_con_f}. Normally, the shortest path problem with hop constraints is also NP-hard. Fortunately, because the weights $(\alpha_{ij} \! + \! \beta_{ij})$ assigned to edges are non-negative, forced by \eqref{pricing_nonnegative}, the hop-constrained shortest path problem can be solved by modifying the Bellman-Ford algorithm \cite{cormen2022introduction}. We denote the shortest path from the MBS to the candidate location $s$, subject to the condition that the path contains no more than $m$ edges, by $u_i^{(m)}$. Initially, we set
\begin{subequations}
\label{BF_method_ini}
\begin{empheq}[left={\empheqlbrace\,}]{align}
& u_0^{(1)} = 0, \label{MBS_ini} \\
& u_i^{(1)} = \alpha_{0i}  +  \beta_{0i}, \; {\rm if} \; (0,i) \in \mathcal{E}, \label{BF_ini1} \\
& u_i^{(1)} = \infty, \; {\rm if} \; (0,i) \notin \mathcal{E}.\label{BF_ini2}
\end{empheq}
\end{subequations}It should be noted that the MBS is indexed by 0. The value of $u_i^{(m)}$ can be updated as follows:
\begin{align}
\label{BF_update}
u_i^{(m+1)} = \min \big\{u_i^{(m)}, \, \min_{(j,i) \in \mathcal{E}}\{u_j^{(m)} + \alpha_{ji}  +  \beta_{ji} \}\big\}.
\end{align} 
Because of the hop constraints, we stop updating the calculation of \eqref{BF_update} when $m \! + \! 1$ is equal to the hop limit $H$. The length of the shortest path from the candidate location $i$ to the MBS, satisfying the maximum hop constraints, is then obtained by the value of $u_i^{(H)}$. The obtained feasible shortest path would be added to the set $\mathcal{P}'$ if $u_i^{(H)} < \eta \! - \! \gamma_i$. Otherwise, there is no feasible route sourcing from $i$ that should be added to $\mathcal{P}'$. According to Lemma 1, the process of column generation would be terminated until the shortest path from all the locations $i \! \in \! \mathcal{V}$ is no less than $(\eta \! - \! \gamma_i)$.

Although the aforementioned column generation method can reduce the problem scale, the achieved result is optimal for the problem (\ref{flow_max_form}-LR). A random rounding method is then proposed to generate an integer solution for \eqref{flow_max_form}. We denote the optimal solution for (\ref{flow_max_form}-LR) by $\{{f^t_p}^*\}$, set the probability of route $p$ being selected as ${f^t_p}^* / D_i^t$, and pick these routes one by one with this probability distribution until the number of candidate locations they passed is equal to $N$, forced by the constraint \eqref{flow_max_RABSnumber_con}. We repeat this random rounding process $k^{max}$ times and choose the best result. 

\subsubsection{Minimize the energy consumption by linear programming}
\label{MinmizeEnergy_LP}

In this subsection, we aim to minimize the energy consumption shown as the denominator of \eqref{energy_efficiency}. The problem can be written from \eqref{formulation} as
\begin{subequations}
\label{energy_min_form}
\begin{align}
\; & \max_{\{y_{(i,j)}^{(t-1,t)}\}} \sum_{(i,j) \in \mathcal{E}} E^{\textrm{fly}}_{(i,j)}y_{(i,j)}^{(t-1,t)} + NT(E^t + E^g) \label{energy_min_obj} \\
s.t.
\;& \eqref{degree_con},\, \eqref{yvar_bin}. \label{energy_mind_con} 
\end{align}
\end{subequations}
Although \eqref{energy_min_form} is still an integer linear programming, the following Lemma 2 proves that the optimal solution can be achieved by solving its linear relaxation.

\textit{\textbf{Lemma 2:}}  The set of constraints \eqref{degree_con} is totally unimodular. 

\textit{Proof:} Please refer to the proposition 2.6 in the section III.1.2 of \cite{wolsey1999integer}

Because the Lemma 2 shows that the problem \eqref{energy_min_form} includes a set of totally unimodular constraints and binary constraints, the optimal solution can be obtained by solving its linear relaxation \cite{wolsey1999integer}, i.e., relax the constraints \eqref{yvar_bin} to $0 \leq y_{(i,j)}^{(t-1,t)} \leq 1 $. 

\vspace{-0.2cm}
\section{Numerical Investigations}
\label{NumericalResults}

\begin{table}[!t]
\centering
\caption{Summary Of Notations}
\label{Notation}
\begin{tabular}{p{5.2 cm}|l}
\hline
\textbf{Parameters} & \textbf{Value}\\
\hline
Carrier frequency and available bandwidth $B$ & 73 GHz, 200 MHz \\
Path loss value for LoS channel & Refer to \cite{semiari2017inter} \\
Maximum spectral efficiency $\textsf{SE}_{\textrm{max}}$ & 4.8 bps/Hz \cite{mcmenamy2019hop}\\
Flying velocity and propulsion power & 18 m/s, 162 W \cite{zeng2019energy} \\
Transmission and grasping power & 10 W, 10 W \cite{liao2023optimal} \\
Epoch duration $\eta$ & 1 hour \\
\hline
\end{tabular}
\vspace{-0.4cm}
\end{table}

The results presented in this section are simulated on a 300$\times$300 $\rm m^2$ Manhattan-type map as depicted in Fig. \ref{Toy_example}, where there are 9 square buildings with the size 90$\times$90 $\rm m^2$ and the road width is 15 m. Lampposts are distributed on the roadside that are used for RABSs grasping, as shown in Fig. \ref{Toy_example}. Accordingly, there are 39 candidate locations and 1 MBS. The parameters related to the communication and power are summarized in Table \ref{Notation}. The traffic distribution in spatial and temporal domians could be found in \cite{wang2015approach}.

\begin{figure}[!t]
\setlength{\abovecaptionskip}{-0.1cm}
\centering
\includegraphics[width=.65\linewidth]{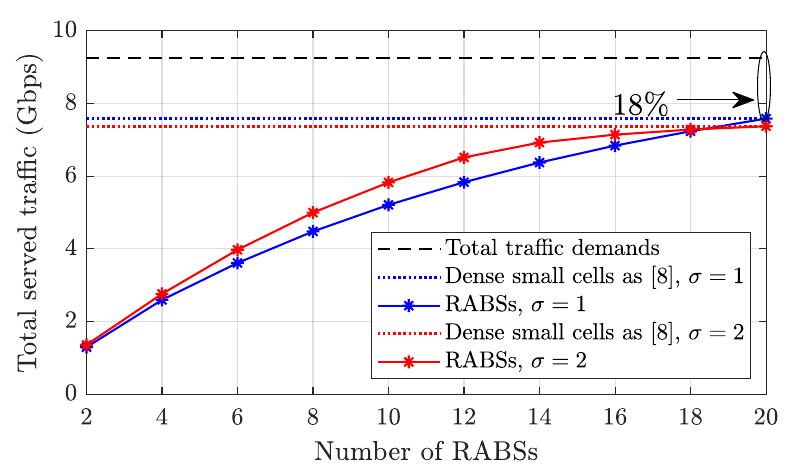}
\caption{Satisfied traffic demand versus the number of RABSs when $H=3$.}
\label{diff_RABS}
\vspace{-0.6cm}
\end{figure}

Fig. \ref{diff_RABS} shows the satisfied traffic demand versus the number of RABSs ranging from 2 to 20 at a certain epoch with different deviation of the traffic distribution, denoted by $\sigma$. Firstly, comparing the served traffic volume versus the total traffic demands, it is shown that deploying only a small group of RABSs can satisfy a large portion of traffic demands. For instance, as marked in Fig. \ref{diff_RABS}, selecting 51\% candidate locations, i.e., 20 out of the total 39 locations, to deploy RABSs can meet nearly 82\% of the total traffic demand. Secondly, the dotted lines represent the densely deployed small cells studied in \cite{mcmenamy2019hop}, that is, deploy small base stations at each candidate location and only determine the route formation. It can be seen that even this dense network cannot satisfy all the traffic demands because of the channel capacity and hop constraints. Also, deploying 20 RABSs can achieve almost the same performance with the dense topology, i.e., using 39 fixed small cells, thanks to their mobility.  

\begin{figure}[!t]
\setlength{\abovecaptionskip}{-0.1cm}
\centering
\includegraphics[width=.65\linewidth]{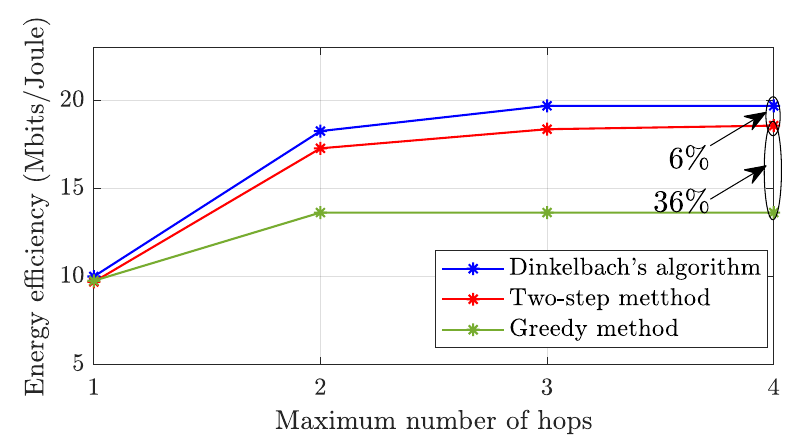}
\caption{Energy efficiency versus the maximum allowed hops when $N=10$.}
\label{diff_hops}
\vspace{-0.4cm}
\end{figure}

Fig. \ref{diff_hops} presents the energy efficiency versus the maximum allowed number of hops solved by different methods. Observe that the energy efficiency grows with the increase of $H$ when $H \! \leq \! 3$. The reason is that a larger $H$ brings a wider coverage to the multi-hop networks so that the candidate locations having high traffic demands but far from MBS have the opportunity to offload data to MBS via multi-hop routes. This growth trend reaches a plateau when $N \! \geq \! 3$ because RABSs can freely choose the candidate locations with high traffic demand and always bakchaul to MBS when $H$ is large enough. Moreover, the proposed two-step method is compared with Dinkelbach's algorithm \cite{you2009dinkelbach} and the greedy method in Fig. \ref{diff_hops}. It should be noted that although the Dinkelbach's algorithm can solve \eqref{formulation} optimally, it requires solving a group of NP-hard integer linear programming problems in each iteration, leading to lengthy computational times when dealing with large-scale problems. In contrast, the proposed two-step method only needs to solve linear programming problems which scale is reduced by the column generation technique in each step, thus it could improve the computational efficiency at the cost of a 6\% optimality gap. Additionally, we employ the greedy method as a benchmark, i.e., at each epoch, it greedily searches the candidate location with the highest traffic volume, calculates the shortest path from the MBS, and adds this shortest path to the solution if it satisfies the hop constraints $H$, otherwise continues to search for the second-best candidate location. Repeat this process until the requirements for RABSs' quantities are met. It can be observed in Fig. \ref{diff_hops} that the two-step method can achieve a higher-quality solution than the greedy method, e.g., exceeding it by 36\% when $H=4$. Therefore, the proposed two-step method achieves a favorable trade-off between high solution quality and low computational complexity.

As mentioned in Section \ref{ColumnGeneration}, the main motivation of employing the column generation is to overcome the exponential growth of the number of feasible routes $|\mathcal{P}|$, shown as the second column in Table \ref{scale_reducation}. The results in Table \ref{scale_reducation} show the fact that the column generation can reduce the problem scale significantly. Numerically, when $H = 5$, there are only 273 routes activated in the restricted master problem \eqref{master_form}, comparing to the number of total feasible routes which is 40128. In other words, the problem scale is reduced by 99\%. Therefore, the column generation approach can solve the problem (\ref{flow_max_form}-LR) to optimality with a restricted subset $\mathcal{P}'$ so that it can improve the efficiency for solving large-scale problems. 

\begin{table}[!t]
\centering
\caption{Problem scale reduced by column generation}
\label{scale_reducation}
\begin{tabular}{|p{1 cm}|p{1.5 cm}|p{1.5 cm}|p{2 cm}|}
\hline
$H$ &\textbf{$|\mathcal{P}|$} & \textbf{$|\mathcal{P}'|$} & \textbf{Scale reduction} \\
\hline
1 & 12 & 12 & 0\% \\
2 & 108 & 66 & 39\% \\
3 & 816 & 177 & 78\%\\
4 & 5802 & 232 & 96\%\\
5 & 40128 & 273 & 99\%\\
\hline
\end{tabular}
\vspace{-0.4cm}
\end{table}

\vspace{-0.2cm}
\section{Conclusions}
\label{Conclusion}
In this letter a novel optimization framework is proposed for network densification via a swarm of aerial robotic base stations (RABSs) that construct an efficient mmWave  network topology for backhauling. To this end, a mixed-integer linear fractional programming (MILFP) problem is formulated and solved by a two-stage method that involves the column generation, Bellman-Ford algorithm and linear programming. Numerical investigations reveal that compared to fixed small cells, only half of RABSs are required to cover the same volume of traffic demand by being able to follow the spatio-temporal traffic dynamics. Hence, RABSs could propel efficient network densification by offering a dramatic reduction on the number of network elements (i.e., small cells) that are required to serve a given traffic demand. Future extensions of this letter may consider the power allocation strategy and integrated access and backhaul technology to further enhance the system performance.
\vspace{-0.4cm}
\bibliographystyle{IEEEtran}
\bibliography{IEEEabrv,reference} 

\end{document}